\documentclass[twocolumn, showpacs, preprintnumbers, superscriptaddress, aps, prl, floatfix, tightenlines]{revtex4-1}  

\usepackage{CJKutf8}
\usepackage{latexsym}
\usepackage{amstext}
\usepackage{amsfonts}
\usepackage{dsfont}
\usepackage{color}
\usepackage{amssymb}
\usepackage{amsmath}
\usepackage{relsize}
\usepackage{amsxtra}
\usepackage{verbatim}
\usepackage{hyperref}
\usepackage{graphicx}
\usepackage{subfigure}

\AtBeginDvi{\input{zhwinfonts}}
\begin{document}
\begin{CJK*}{UTF8}{zhkai}

\title{Incommensurate antiferromagnetic order in the manifoldly-frustrated \\ 
SrTb$_2$O$_4$ with transition temperature up to 4.28 K}

\author{Hai-Feng Li} 
\email{h.li@fz-juelich.de}
\affiliation{J$\ddot{u}$lich Centre for Neutron Science JCNS, Forschungszentrum J$\ddot{u}$lich GmbH, Outstation at Institut Laue-Langevin, Bo$\hat{\imath}$te Postale 156, F-38042 Grenoble Cedex 9, France}
\affiliation{Institut f$\ddot{u}$r Kristallographie der RWTH Aachen University, D-52056 Aachen, Germany}
\author{Cong Zhang}
\affiliation{Institut f$\ddot{u}$r Kristallographie der RWTH Aachen University, D-52056 Aachen, Germany}
\author{Anatoliy Senyshyn}
\affiliation{Forschungsneutronenquelle Heinz Maier-Leibnitz FRM-II, Technische Universit$\ddot{\rm a}$t M$\ddot{\rm u}$nchen, Lichtenbergstrasse 1, D-85747 Garching bei M$\ddot{\rm u}$nchen, Germany}
\author{Andrew Wildes}
\affiliation{Institut Laue-Langevin, Bo$\hat{\imath}$te Postale 156, F-38042 Grenoble Cedex 9, France}
\author{Karin Schmalzl} 
\affiliation{J$\ddot{u}$lich Centre for Neutron Science JCNS, Forschungszentrum J$\ddot{u}$lich GmbH, Outstation at Institut Laue-Langevin, Bo$\hat{\imath}$te Postale 156, F-38042 Grenoble Cedex 9, France}
\author{Wolfgang Schmidt}
\affiliation{J$\ddot{u}$lich Centre for Neutron Science JCNS, Forschungszentrum J$\ddot{u}$lich GmbH, Outstation at Institut Laue-Langevin, Bo$\hat{\imath}$te Postale 156, F-38042 Grenoble Cedex 9, France}
\author{Martin Boehm}
\affiliation{Institut Laue-Langevin, Bo$\hat{\imath}$te Postale 156, F-38042 Grenoble Cedex 9, France}
\author{Eric Ressouche}
\affiliation{SPSMS, UMR-E 9001, CEA-INAC/UJF-Grenoble 1, MDN, 17 rue des Martyrs, F-38054 Grenoble Cedex 9, France}
\author{Binyang Hou}
\affiliation{European Synchrotron Radiation Facility, Bo$\hat{\imath}$te Postale 220, F-38043 Grenoble Cedex, France}
\author{Paul Meuffels}
\affiliation{Peter Gr$\ddot{u}$nberg Institut PGI and JARA-FIT, Forschungszentrum J$\ddot{u}$lich GmbH, D-52425 J$\ddot{u}$lich, Germany}
\author{Georg Roth}
\affiliation{Institut f$\ddot{u}$r Kristallographie der RWTH Aachen University, D-52056 Aachen, Germany}
\author{Thomas Br$\ddot{\texttt{u}}$ckel}
\affiliation{J$\ddot{u}$lich Centre for Neutron Science JCNS and Peter Gr$\ddot{u}$nberg Institut PGI, JARA-FIT, Forschungszentrum J$\ddot{u}$lich GmbH, D-52425 J$\ddot{u}$lich, Germany}

\date{\today}

\begin{abstract}

The N$\acute{\rm e}$el temperature of the new frustrated family of Sr\emph{RE}$_2$O$_4$ (\emph{RE} = rare earth) compounds is yet limited to $\sim$ 0.9 K, which more or less hampers a complete understanding of the relevant magnetic frustrations and spin interactions. Here we report on a new frustrated member to the family, SrTb$_2$O$_4$ with a record $T_{\rm N}$ = 4.28(2) K, and an experimental study of the magnetic interacting and frustrating mechanisms by polarized and unpolarized neutron scattering. The compound SrTb$_2$O$_4$ displays an incommensurate antiferromagnetic (AFM) order with a transverse wave vector \textbf{Q}$^{\rm 0.5 K}_{\rm AFM}$ = (0.5924(1), 0.0059(1), 0) albeit with partially-ordered moments, 1.92(6) $\mu_{\rm B}$ at 0.5 K, stemming from only one of the two inequivalent Tb sites mainly by virtue of their different octahedral distortions. The localized moments are confined to the \emph{bc} plane, 11.9(66)$^\circ$ away from the \emph{b} axis probably by single-ion anisotropy. We reveal that this AFM order is dominated mainly by dipole-dipole interactions and disclose that the octahedral distortion, nearest-neighbour (NN) ferromagnetic (FM) arrangement, different next NN FM and AFM configurations, and in-plane anisotropic spin correlations are vital to the magnetic structure and associated multiple frustrations. The discovery of the thus far highest AFM transition temperature renders SrTb$_2$O$_4$ a new friendly frustrated platform in the family for exploring the nature of magnetic interactions and frustrations.

\end{abstract}


\maketitle
\end{CJK*}


\section{I. Introduction}

Revealing the magnetic coupling mechanism is often a critical step towards understanding the role of magnetism in intriguing phenomena such as colossal magnetoresistance (CMR), high $T_{\rm C}$ superconductivity, multiferroicity or frustration in correlated electron materials \cite{Santen1950, Bednorz1986, Fiebig2002, Diep2004, Lacroix2011}. By way of example, the indirect double- and super-exchange interactions were successfully elaborated in qualitatively explaining the CMR effect and associated magnetic orders based only on the spin and charge degrees of freedom \cite{Santen1950}. In 4\emph{f}-based insulators, the indirect oscillating interaction \cite{Diep2004, Lacroix2011, Li2012, Feng2013} between pairs of localized 4\emph{f} moments via the intermediary of valence electrons is blocked. Therefore, possible super-, dipole-dipole and multipolar, and Dzyaloshinsky-Moriya (DM) exchange interactions are primarily responsible for potential magnetic ordering \cite{Diep2004}. Without detailed knowledge of the structural and magnetic parameters, it is hard to uniquely determine which interaction acts as the major exchange mechanism \cite{Xiao2010}. In this case, the origins of the related incommensurable spin structures become elusive \cite{Diep2004}. In addition, the competition between spin-orbital coupling and crystal electric field (CEF) at low temperatures largely affects the highly-degenerate Hund's rule ground state, and besides the anisotropic dipolar and DM interactions, determine the magnitude of the magnetic anisotropy \cite{Tian2010, Li2014-2}. This anisotropy strongly influences the degree of magnetic frustration. Sometimes, it may disorder or even quench potential magnetic moments, leading to a virtually nonmagnetic ground state \cite{Li2014-1}.

\begin{figure*}[!ht]
\centering \includegraphics[width = 0.78\textwidth] {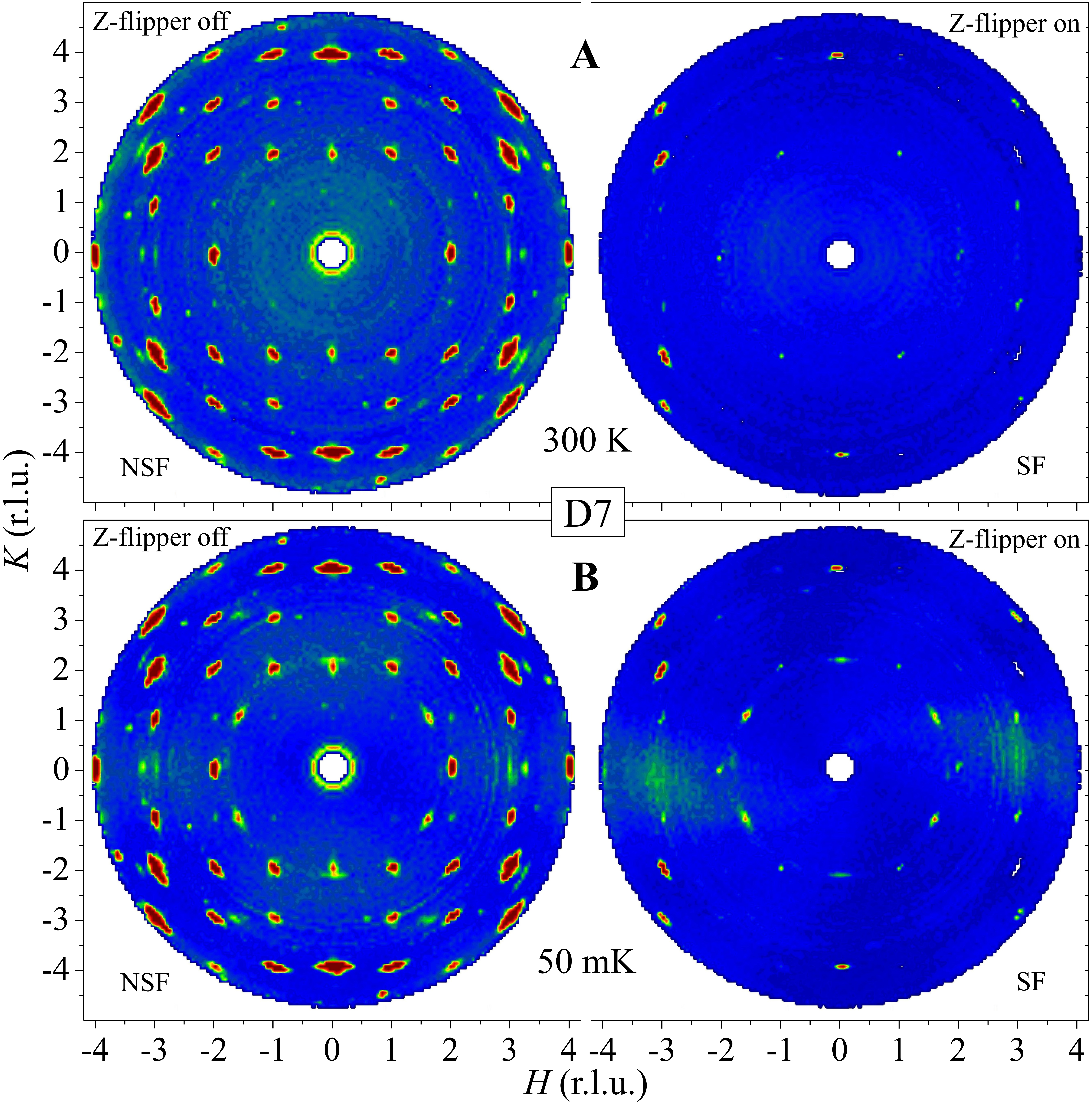}
\caption{
\textbf{Polarization analysis data measured using D7 (ILL).}
\textbf{(A)} At 300 K. \textbf{(B)} At 50 mK. The NSF (i.e., Z-flipper off, left panel) and SF (i.e., Z-flipper on, right panel) channels are shown with the same colour code for intensity. The non-perfect polarization involuntarily leads to the presence of some nuclear Bragg peaks, e.g. (0, $\pm$4, 0), in the SF channel at both temperatures. The horizontal bar-shaped neutron-scattering intensities around (0, $\pm$2.15, 0) and the extremely-broad diffuse scattering around ($\pm$3, 0, 0) shown in \textbf{(B)} may correspond to some short-range magnetic components. It is pointed out that similar diffuse magnetic scattering also appears in the SrHo$_2$O$_4$ and SrEr$_2$O$_4$ single crystals \cite{Ghosh2011, Petrenko2008, Hayes2012}.
}
\label{Fig1}
\end{figure*}

\begin{figure}[!ht]
\centering \includegraphics[width = 0.48\textwidth] {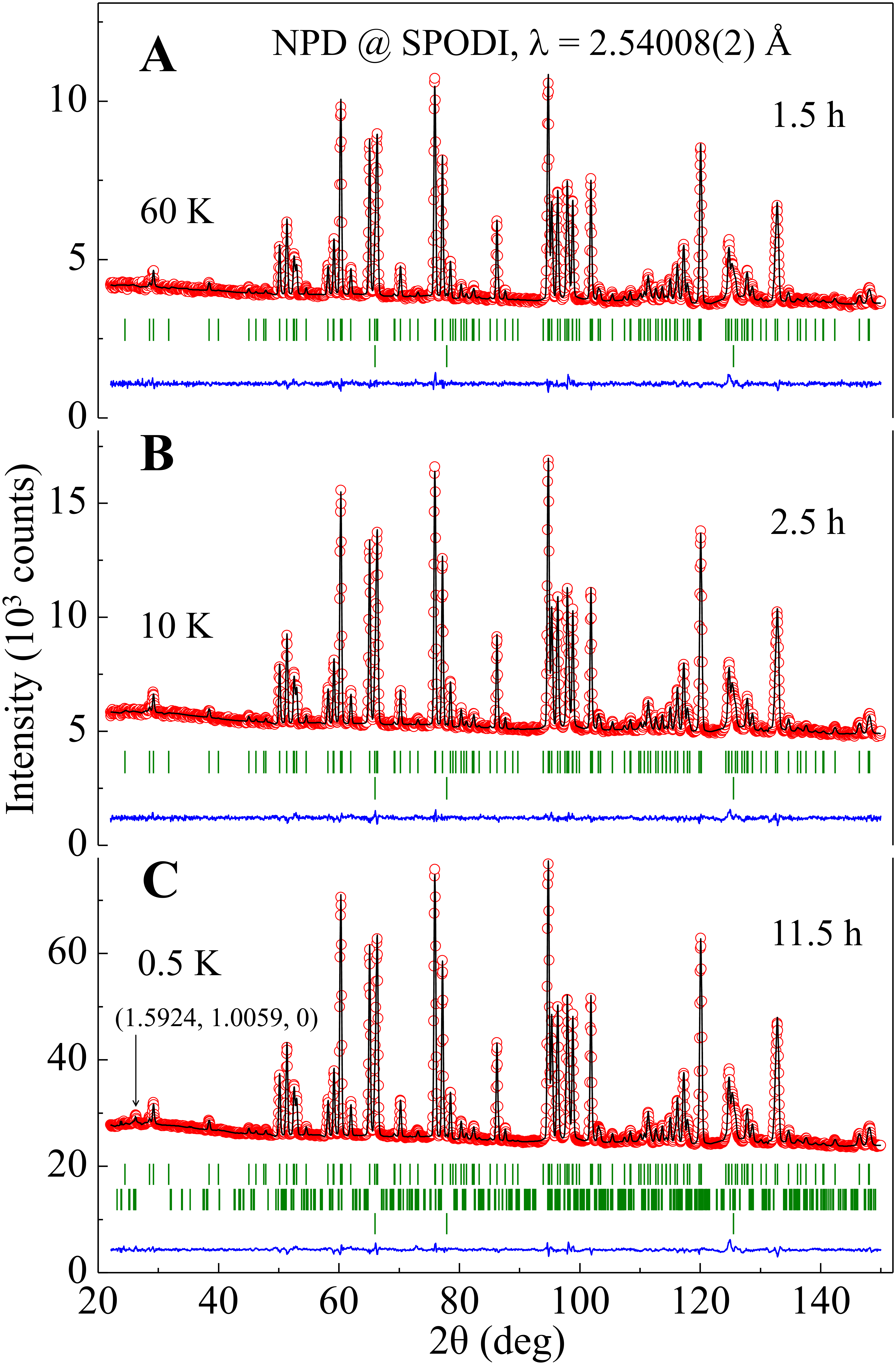}
\caption{
\textbf{Observed (circles) and calculated (solid lines) NPD patterns from the study using SPODI (FRM-II).}
\textbf{(A)} At 60 K with counting time $\sim$ 1.5 h. \textbf{(B)} At 10 K with counting time $\sim$ 2.5 h. (\textbf{C}) At 0.5 K with counting time $\sim$ 11.5 h. The vertical bars mark the positions of nuclear and magnetic Bragg reflections of SrTb$_2$O$_4$ as well as the Al nuclear Bragg peaks (from sample environment), respectively. The lower curves represent the difference between observed and calculated patterns. Here appears no obvious diffuse magnetic scattering as observed in polycrystalline Sr$RE_2$O$_4$ (\emph{RE} = Ho, Er, Dy) compounds in reference \cite{Karunadasa2005}.
}
\label{Fig2}
\end{figure}

\begin{figure*}[!ht]
\centering \includegraphics[width = 0.78\textwidth] {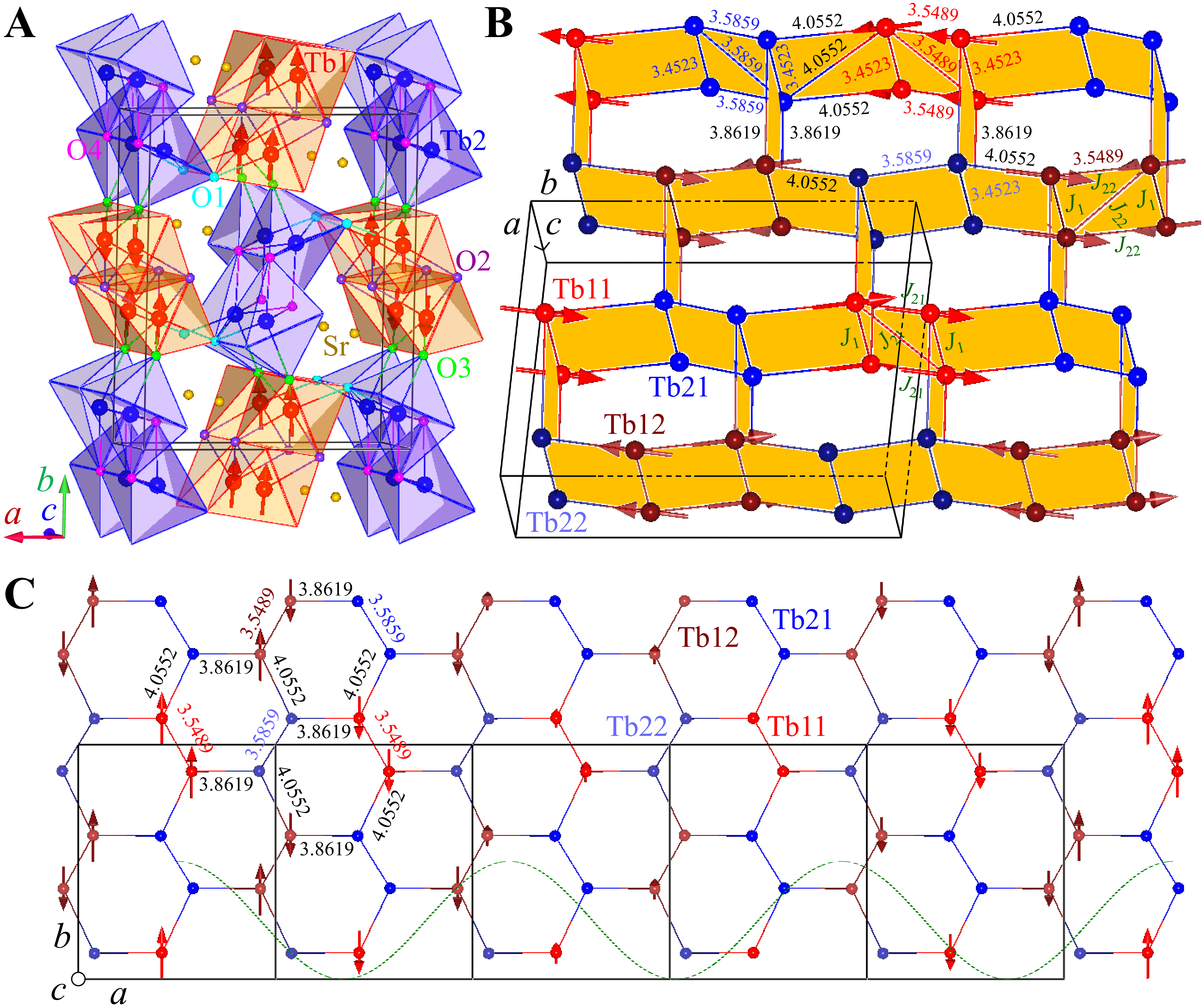}
\caption{\textbf{Crystal and magnetic structures of SrTb$_2$O$_4$.}
\textbf{(A)} As refined with the SPODI (FRM-II) data measured at 0.5 K. \textbf{(B)} Detailed Tb-Tb bond lengths within the bent Tb$_6$ honeycombs. $J_1$ and $J_{21}$/$J_{22}$ represent the nearest-neighbor (NN) and next-NN (NNN) magnetic couplings. \textbf{(C)} Projection of the bent Tb$_6$ honeycombs to the \emph{ab} plane. The cosine-curve (dashed line) beginning from the center of the first unit-cell in left schematically shows the incommensurable spin modulation along the \emph{a} axis. In \textbf{(A-C)}, the arrows drawn through the Tb1 ions represent the Tb1 partially-ordered moments, and the connected lines represent the crystallographic unit cell(s). To clearly show the magnetic frustration, Tb1 and Tb2 sites in \textbf{(A)} are transferred into Tb11 and Tb12 as well as Tb21 and Tb22 sites, respectively, as shown in \textbf{(B)} and \textbf{(C)} by the irreducible representation analysis to the \emph{P}-1 symmetry \cite{Rodriguez-Carvajal1993}.
}
\label{Fig3}
\end{figure*}

\begin{figure}[!ht]
\centering \includegraphics[width = 0.48\textwidth] {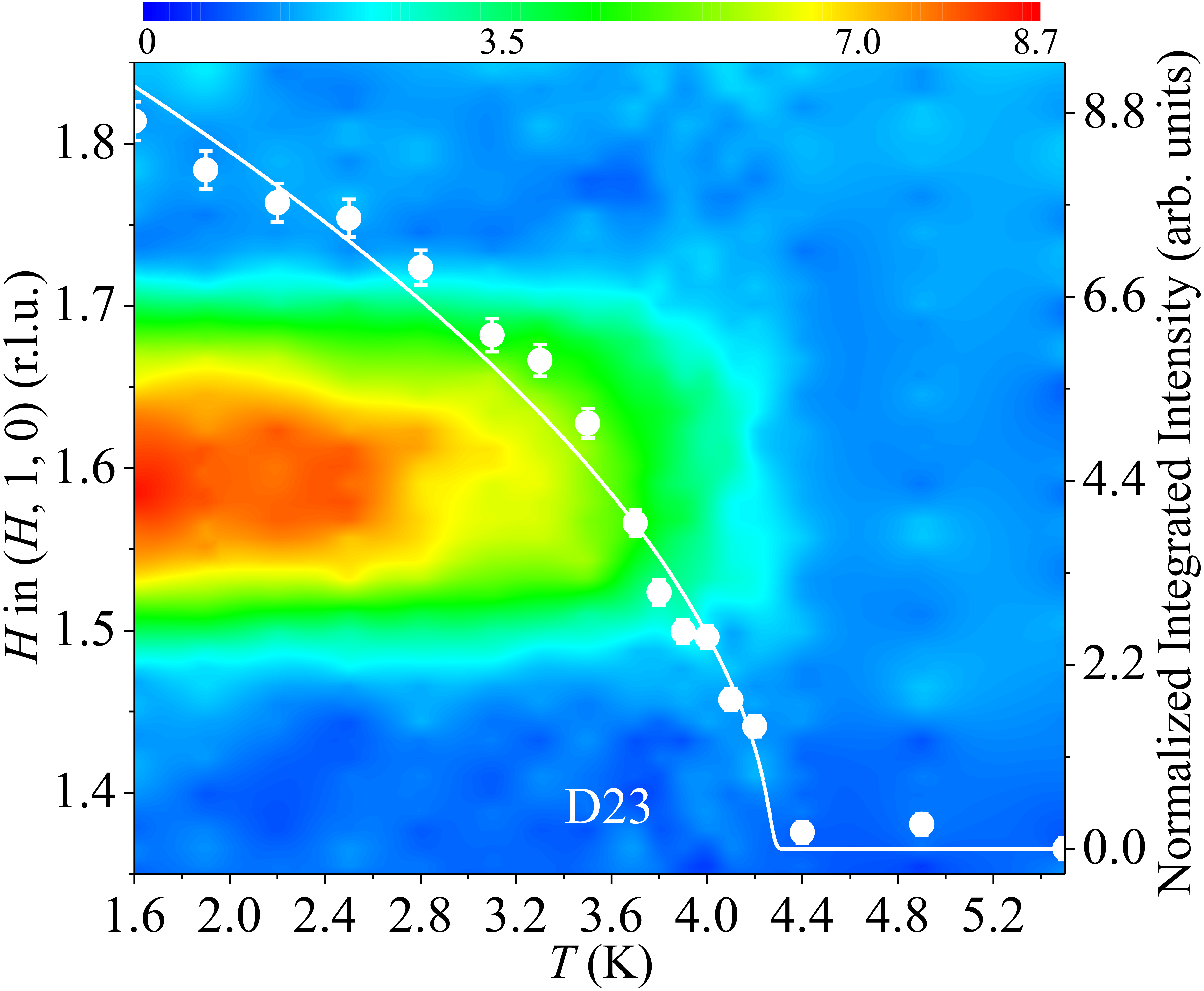}
\caption{\textbf{Temperature-dependent ordering parameter of the AFM transition.}
Temperature dependent \emph{H}-scans around the AFM (1.6, 1, 0) Bragg peak (left ordinate) performed at D23 (ILL) and the corresponding integrated intensities (circles) (right ordinate). The solid line is a fit with the power law as described in the text. The error bars are statistical errors.
}
\label{Fig4}
\end{figure}

\begin{figure}[!ht]
\centering \includegraphics[width = 0.48\textwidth] {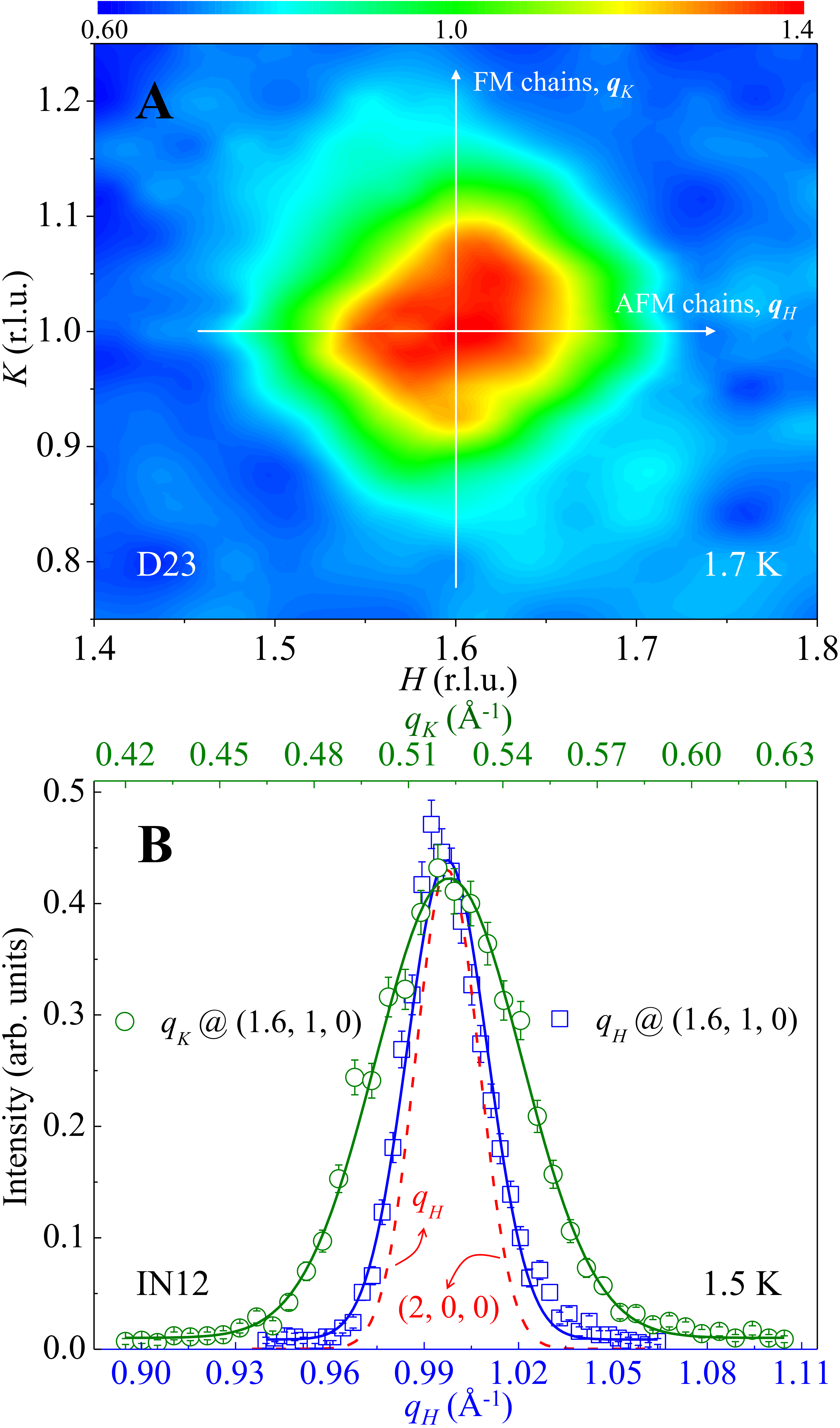}
\caption{\textbf{Anisotropic in-plane spin correlations of SrTb$_2$O$_4$.}
\textbf{(A)} Unpolarized neutron data around the AFM (1.6, 1, 0) Bragg reflection measured at 1.7 K using D23 (ILL). The two perpendicular arrows point out the directions of $\textbf{\emph{q}}_H$ (i.e., along the in-plane AFM chains) and $\textbf{\emph{q}}_K$ (i.e., along the in-plane FM chains), respectively. \textbf{(B)} Unpolarized neutron data around the (1.6, 1, 0) (void symbols) and the (2, 0, 0) (dashed line) Bragg peaks measured at 1.5 K using IN12 (ILL). For comparison, the $q_H$ values around the nuclear (2, 0, 0) Bragg peak are reduced by 0.2547 {\AA}$^{-1}$. The solid lines are Gaussian fits to the magnetic data convoluted with the nuclear correlation length as expected \textbf{Q}-resolution.
}
\label{Fig5}
\end{figure}

\begin{figure}[!ht]
\centering \includegraphics[width = 0.48\textwidth] {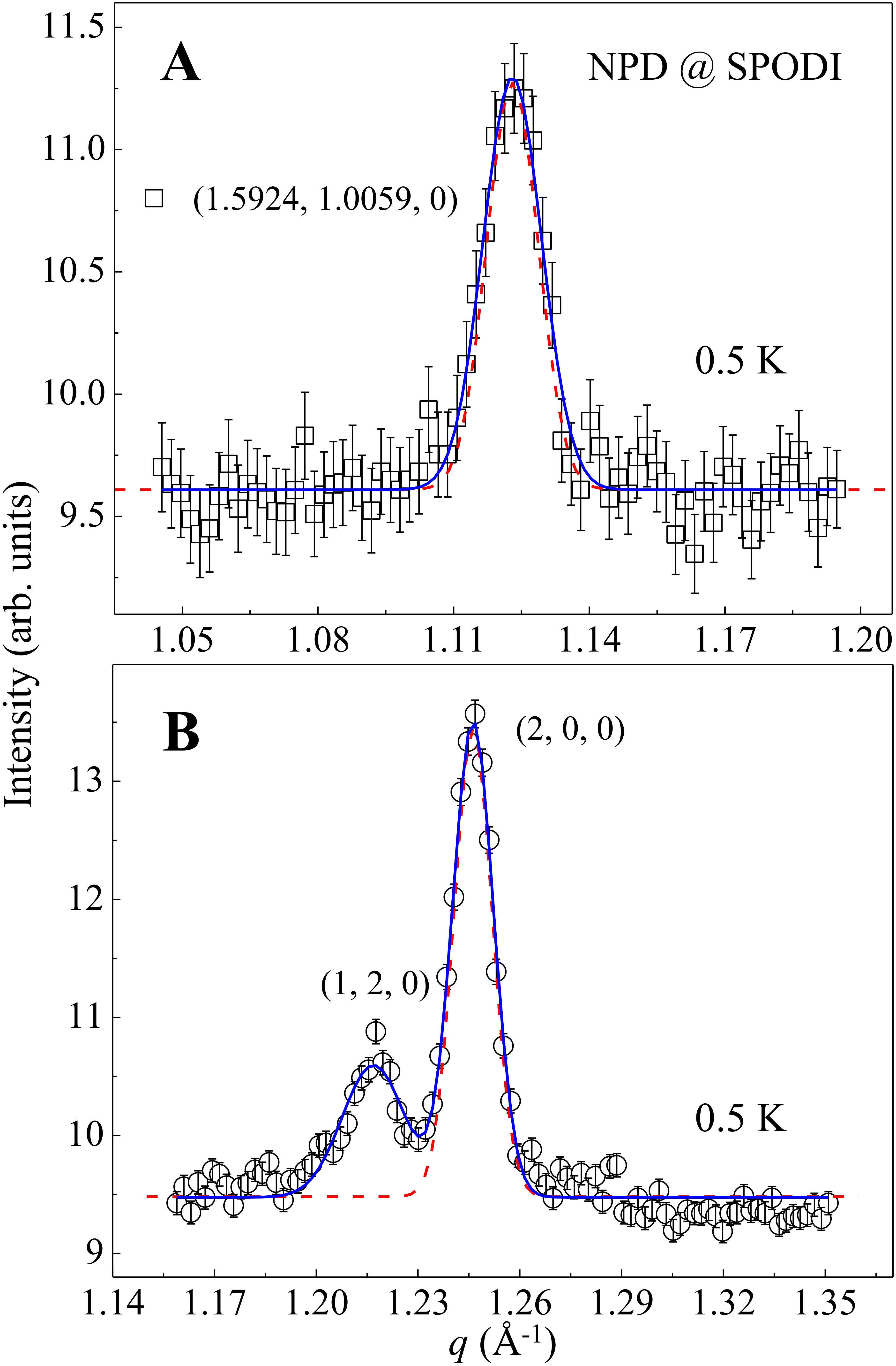}
\caption{\textbf{NPD peaks of the incommensurate AFM (1.5924, 1.0059, 0) and the nuclear Bragg (2, 0, 0) reflections.}
The data were measured at 0.5 K using SPODI (FRM-II) and extracted from \textbf{Figure~\ref{Fig2}C}. \textbf{(A)} NPD data of the incommensurate AFM (1.5924, 1.0059, 0) Bragg reflection (squares), and \textbf{(B)} of the nuclear Bragg (1, 2, 0) and (2, 0, 0) reflections (circles). The solid lines are Gaussian fits convoluted with the corresponding \textbf{Q}-resolutions (dashed lines) \cite{Hoelzel2012}.
}
\label{Fig6}
\end{figure}

\begin{figure*}[!ht]
\centering \includegraphics[width = 0.78\textwidth] {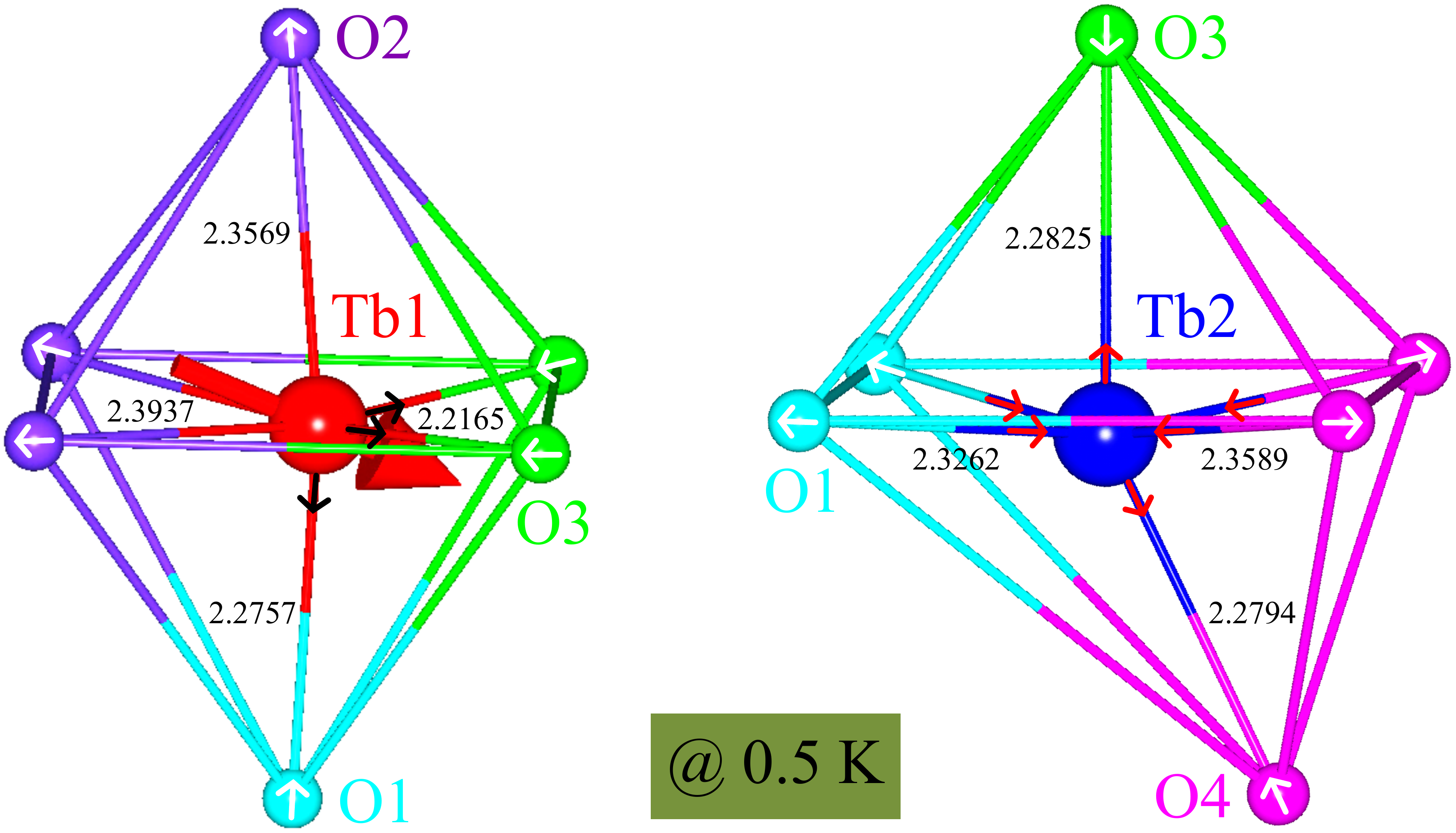}
\caption{\textbf{Local octahedral environment of the two Tb sites.}
The detailed Tb-O bond lengths within the TbO$_6$ octahedra as refined from the SPODI (FRM-II) data collected at 0.5 K. The big arrow drawn through the Tb1 ion represents the Tb1 moment. The small arrows sitting on the O ions or the Tb-O bonds schematically show the deduced octahedral distortion modes as displayed.
}
\label{Fig7}
\end{figure*}

\begin{figure}[!ht]
\centering \includegraphics[width = 0.48\textwidth] {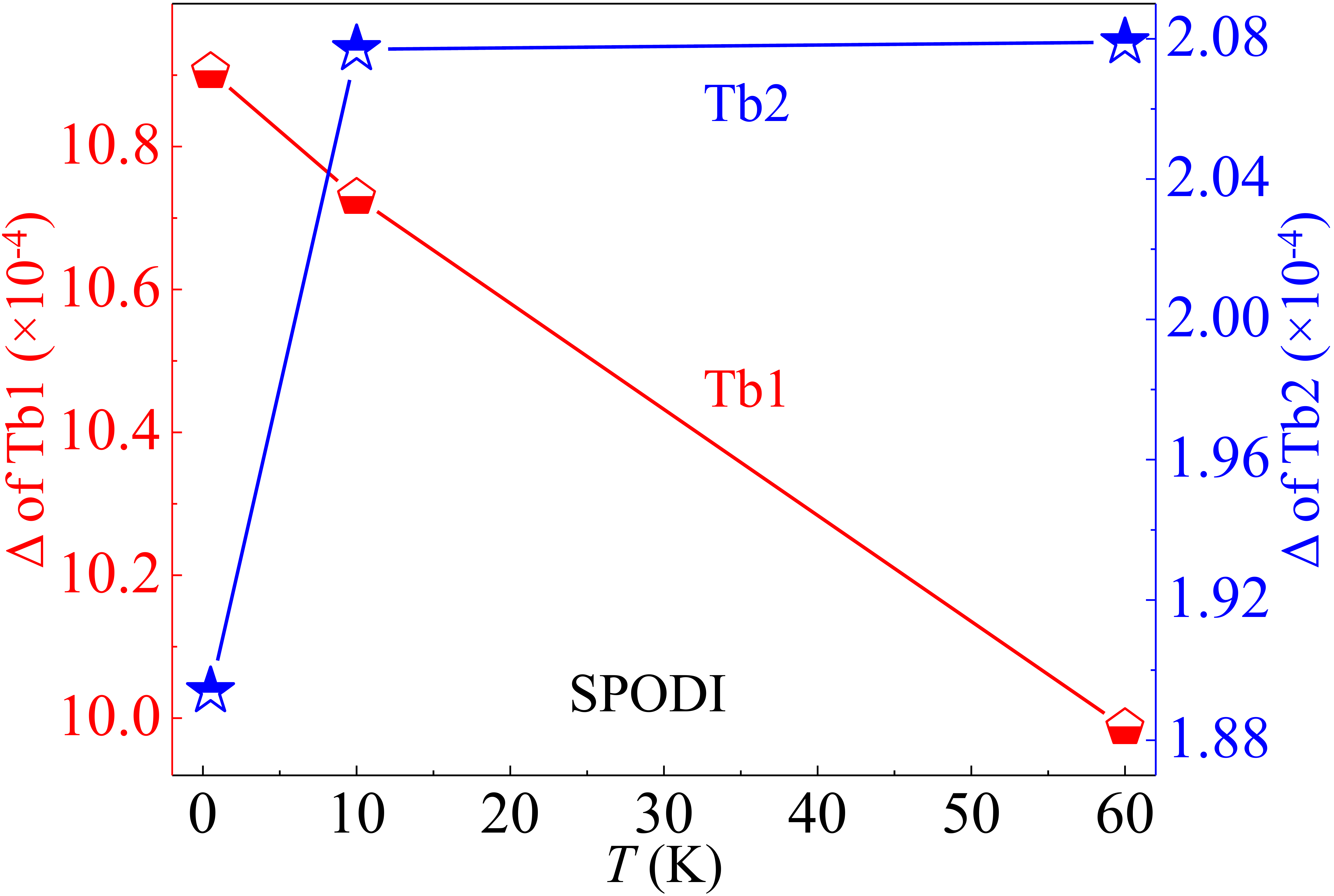}
\caption{\textbf{Octahedral distortions of the Tb1 and Tb2 sites.}
Octahedral distortion parameter $\Delta$ of the Tb1 (left ordinate) and Tb2 (right ordinate) sites as refined from the SPODI (FRM-II) data.
}
\label{Fig8}
\end{figure}

Magnetic frustration can lead to novel quantum states such as spin liquid, spin ice, cooperative paramagnetism or the magnetic Coulomb phase based on magnetic monopole excitations, providing an excellent testing ground for theories \cite{Diep2004, Morris2009, Gardner1999, Lake2000, Braun2005, Harris1997, Bramwell2001, Lee2002, Ramirez2003, Moessner2006, Han2008, Gingras2009, Evgeny2013, Picozzi2014, Morgan2013}. A Monte Carlo simulation indicates that the observed diffuse scattering in SrEr$_2$O$_4$ originates from a ladder of Er triangles \cite{Hayes2011}. A computation of the crystal-field levels demonstrates site-dependent anisotropic single-ion magnetism in the compounds of SrHo$_2$O$_4$ and SrDy$_2$O$_4$ \cite{Fennell2014}. Lanthanide-based magnetic compounds, e.g. edge-sharing tetrahedra, corner-sharing spinels, or triangular Kagom$\acute{\rm e}$ and pyrochlore lattices, often show anomalous magnetic properties due to geometric frustration \cite{Kawamura1998, Ramirez1990, Binder1980, Petrenko2014}. The family of Sr\emph{RE}$_2$O$_4$ (\emph{RE} = Y, Gd, Ho, Yb) compounds was first synthesized in 1967 \cite{Barry1967}. Recently, a study on polycrystalline Sr\emph{RE}$_2$O$_4$ (\emph{RE} = Gd, Dy, Ho, Er, Tm, Yb) samples demonstrates that they adopt the orthorhombic structure \cite{Pepin1981} with a geometric frustration for the magnetic ions revealed by the existence of magnetic short-range orders down to $\sim$ 1.5 K \cite{Karunadasa2005}. Subsequently, single crystals of Sr\emph{RE}$_2$O$_4$ (\emph{RE} = Y, Lu, Dy, Ho, Er) were successfully grown \cite{Balakrioshnan2009}. Single-crystal neutron-scattering studies on Sr\emph{RE}$_2$O$_4$ (\emph{RE} = Ho, Er, Yb) compounds with respective antiferromagnetic (AFM) transition temperatures at 0.62, 0.73, and 0.9 K were reported \cite{Ghosh2011, Wen2014, Petrenko2008, Quintero2012, Hayes2012}, generally confirming that there exists a coexistence of long- and short-range magnetic orders. It is pointed out that for the case of SrHo$_2$O$_4$, Young, et al. \cite{Young2013} observed only a short-range spin order inconsistent with other reports \cite{Ghosh2011, Wen2014, Hayes2012}. Further experimental tests would be necessary to address this discrepancy. Since the adopted orthorhombic structure accommodates two \emph{RE} sites (\emph{RE1} and \emph{RE2}), it is hard to derive the crystallographic origins of the two types of spin ordering. In addition, single-crystal SrDy$_2$O$_4$ displays only weak diffuse magnetic scattering which persists down to $\sim$ 20 mK \cite{Cheffings2013}. The low transition temperatures of the magnetic orders, to some extent, prevent a complete understanding of the nature of magnetic interactions and frustrations in the family. To overcome these problems and address the relevant interesting physics necessitate a search in the Sr\emph{RE}$_2$O$_4$ (\emph{RE} = rare earth) family for a new compound that displays a higher N$\acute{\texttt{e}}$el temperature, thus permitting a technically easier study of the two coupling mechanisms.

In this study, we report on a new frustrated member to the family of Sr\emph{RE}$_2$O$_4$, namely SrTb$_2$O$_4$, which has not been studied yet by neutron scattering. The single-crystal SrTb$_2$O$_4$ displays a long-range magnetic order relative to the underlying lattice. The noncollinear incommensurate AFM structure forms at $T_{\rm N}$ = 4.28(2) K upon cooling. The synthesis of SrTb$_2$O$_4$ with the highest N$\acute{\texttt{e}}$el temperature in the family opens up an easier route to elucidate the magnetic coupling and frustrating mechanisms. By polarized and unpolarized neutron scattering we uniquely determine the detailed structural and magnetic parameters to understand the magnetism in SrTb$_2$O$_4$.

\section{II. Experimental}

Polycrystalline samples of SrTb$_2$O$_4$ were synthesized from stoichiometric mixtures of SrCO$_3$ (99.99\%) and Tb$_4$O$_7$ (99.99\%) compounds by standard solid-state reaction \cite{Li2008}. Both raw materials were preheated at 800$^\circ$ for 12 h and weighted at $\sim$ 200$^\circ$. The mixed and milled raw materials were calcined twice at 1473 and 1573 K for 48 h each in air in order to perform decarbonization and prereaction. The resulting powder was pressed into cylindrical rods with an isostatical pressure of $\sim$ 78 MPa. The rods were sintered two times at 1573 and 1673 K for 48 h at each temperature in air. After each round of the isostatic pressing and subsequent firing, the product was reground and ball-remilled, which results in a dense and homogenous sample and ensures a complete chemical reaction. The single crystal of SrTb$_2$O$_4$ was grown by optical floating-zone method with an atmosphere of $\sim$ 98\% Ar and $\sim$ 2\% O$_2$. The growing speed is $\sim$ 4 mm/h with rotations of the feed and seed rods at +32 and -28 rpm, respectively. The phase purity of the polycrystalline and single-crystalline samples was checked by in-house X-ray powder diffraction. The electrical resistivity of a bar-shaped single crystal by standard dc four-probe technique was measured on a commercial physical property measurement system.

High-resolution neutron powder diffraction (NPD) patterns were collected with a pulverized SrTb$_2$O$_4$ single crystal ($\sim$ 5 g) mounted in a $^3$He insert on the structure powder diffractometer (SPODI) \cite{Hoelzel2012} with constant wavelength $\lambda$ = 2.54008(2) {\AA} at the FRM-II research reactor in Garching, Germany.

The SrTb$_2$O$_4$ single crystal ($\sim$ 2.2 g) for the neutron-scattering studies was oriented in the (\emph{H}, \emph{K}, 0) scattering plane with the neutron Laue diffractometer OrientExpress \cite{Ouladdiaf2006} and the IN3 thermal triple-axis spectrometer at the Institut Laue-Langevin (ILL), Grenoble, France. The mosaic of this single crystal is 0.494(5)$^\texttt{o}$ full width at half maximum (FWHM) for the nuclear (2, 0, 0) Bragg reflection at 1.5 K.
Longitudinal \emph{XYZ} neutron polarization analysis was carried out on the D7 (ILL) diffractometer \cite{Stewart2009} with a dilution fridge and $\lambda$ = 4.8 {\AA}. Unpolarized elastic neutron-scattering studies were performed at the two-axis D23 diffractometer (ILL) with incident wavelength 1.277 {\AA} and the IN12 (ILL) cold triple-axis spectrometer with fixed final energy of 5.3 meV and the beam collimation set as open-40$'$-sample-60$'$-open.

Here the wave vector \textbf{Q}$_{(HKL)}$ ({\AA}$^{-1}$) = ($\textbf{Q}_H$, $\textbf{Q}_K$, $\textbf{Q}_L$) is defined through (\emph{H}, \emph{K}, \emph{L}) = ($\frac{a}{2\pi}Q_H$, $\frac{b}{2\pi}Q_K$, $\frac{c}{2\pi}Q_L$) quoted in units of r.l.u., where \emph{a}, \emph{b}, and \emph{c} are relevant lattice constants referring to the orthorhombic \cite{Pepin1981} unit cell.

\section{III. Results}

\textbf{Figure~\ref{Fig1}} shows the neutron polarization analysis in the spin-flip (SF, i.e., Z-flipper on) and non-spin-flip (NSF, i.e., Z-flipper off) channels. Compared with the maps at 300 K (\textbf{Figure~\ref{Fig1}A}), it is clear that extra fourfold Bragg peaks around ($\pm$1.6, $\pm$1, 0) appear symmetrically in both SF and NSF reciprocal space maps at 50 mK (\textbf{Figure~\ref{Fig1}B}) due to a long-range magnetic transition. Polarized neutron magnetic scattering depends on the direction of the neutron polarization $\hat{\textbf{P}}$ with respect to the scattering vector $\hat{\textbf{Q}}$, and also the direction of the ordered-moments $\hat{\mu}$. In our case, $\hat{\textbf{P}}$ (Z-component) $\parallel$ \emph{c}-axis \cite{Stewart2009}, and the magnetic Bragg reflections are observed in the (\emph{H}, \emph{K}, 0) plane, i.e., $\hat{\textbf{P}}$ $\perp$ $\hat{\textbf{Q}}$. In this case, the neutron-scattering cross sections of the NSF and SF channels are
\begin{eqnarray}
\label{eq1}
&& \mathlarger{\mathlarger{\mathlarger{(}}}\frac{d\sigma}{d\Omega}\mathlarger{\mathlarger{\mathlarger{)}}}^{^{\texttt{NSF}}}_{_{\texttt{Z-off}}} = \frac{1}{2}\mathlarger{\mathlarger{\mathlarger{(}}}\frac{d\sigma}{d\Omega}\mathlarger{\mathlarger{\mathlarger{)}}}_{_{\texttt{mag}}} + \frac{1}{3}\mathlarger{\mathlarger{\mathlarger{(}}}\frac{d\sigma}{d\Omega}\mathlarger{\mathlarger{\mathlarger{)}}}_{_{\texttt{si}}} + \mathlarger{\mathlarger{\mathlarger{(}}}\frac{d\sigma}{d\Omega}\mathlarger{\mathlarger{\mathlarger{)}}}_{_{\texttt{nuc}}} \text{,} \nonumber \\*
&& \mathlarger{\mathlarger{\mathlarger{(}}}\frac{d\sigma}{d\Omega}\mathlarger{\mathlarger{\mathlarger{)}}}^{^{\texttt{NSF}}}_{_{\texttt{mag}}} \text{\textcolor[rgb]{1.00,1.00,1.00}{ab}} {\propto} \text{\textcolor[rgb]{1.00,1.00,1.00}{ab}} {\langle}{\hat{\mu}}\parallel\hat{\textbf{P}}{\rangle}^2 \text{, and} \\*
\label{eq2}
&& \mathlarger{\mathlarger{\mathlarger{(}}}\frac{d\sigma}{d\Omega}\mathlarger{\mathlarger{\mathlarger{)}}}^{^{\texttt{SF}}}_{_{\texttt{Z-on}}} \text{\textcolor[rgb]{1.00,1.00,1.00}{,}} = \frac{1}{2}\mathlarger{\mathlarger{\mathlarger{(}}}\frac{d\sigma}{d\Omega}\mathlarger{\mathlarger{\mathlarger{)}}}_{_{\texttt{mag}}} + \frac{2}{3}\mathlarger{\mathlarger{\mathlarger{(}}}\frac{d\sigma}{d\Omega}\mathlarger{\mathlarger{\mathlarger{)}}}_{_{\texttt{si}}} \text{,} \nonumber \\*
&& \mathlarger{\mathlarger{\mathlarger{(}}}\frac{d\sigma}{d\Omega}\mathlarger{\mathlarger{\mathlarger{)}}}^{^{\texttt{SF}}}_{_{\texttt{mag}}} \text{\textcolor[rgb]{1.00,1.00,1.00}{ab}} {\propto} \text{\textcolor[rgb]{1.00,1.00,1.00}{ab}} {\langle}{\hat{\mu}}\perp\hat{\textbf{P}} \times \hat{\textbf{Q}}{\rangle}^2 \text{,}
\end{eqnarray}
respectively. The first and the second terms in each equation refer to the magnetic and spin-incoherent scatterings, respectively. The third term in Eq.~(\ref{eq1}) denotes nuclear and isotope incoherent contributions \cite{Stewart2009}. The presence of the incommensurable AFM Bragg peaks in the NSF channel (\textbf{Figure~\ref{Fig1}B}) indicates that one component of $\hat{\mu}$ is parallel to the \emph{c} axis, while their appearances in the SF channel imply a $\hat{\mu}$ component lying in the \emph{ab} plane.

We observe the magnetic Bragg peak only at 0.5 K in our NPD study (\textbf{Figure~\ref{Fig2}}). We thereby refine the AFM wave vector exactly as \textbf{Q}$_{\texttt{AFM}}$ = (0.5924(1), 0.0059(1), 0) by the profile-matching mode \cite{Rodriguez-Carvajal1993} and a total moment $|\hat{\mu}|$ = 1.92(6) $\mu_\texttt{B}$ at the maximum amplitude for the Tb1 ions only with the \emph{b}- and \emph{c}-components equalling to +1.88(8) and +0.40(23) $\mu_\texttt{B}$ (\textbf{Table 1}), respectively. The moment size of the Tb2 site is negligible. \textbf{Figure~\ref{Fig3}} schematically shows the resulting crystal and magnetic structures as well as the structural parameters for the bent Tb$_6$ honeycombs. The temperature dependence of the AFM (1.6, 1, 0) Bragg peak is shown in \textbf{Figure~\ref{Fig4}}. The extracted integrated intensity (\emph{I}) was fit to a power law $I = I_0(1-\frac{T}{T_{\texttt{N}}})^\beta$, which produces a N$\acute{\texttt{e}}$el temperature $T_{\texttt{N}} =$ 4.28(2) K, and a critical exponent $\beta =$ 0.55(2) probably indicative of a second-order type phase transition and possible three-dimensional Heisenberg-like spin interactions \cite{Collins1989}.

We record a reciprocal space map (\textbf{Figure~\ref{Fig5}A}) around the AFM (1.6, 1, 0) Bragg peak at 1.7 K using D23, and the central scans along the $q_H$ and $q_K$ directions (\textbf{Figure~\ref{Fig5}B}) were measured at IN12. In both \textbf{Figures}, the FWHM of the magnetic Bragg peak along the $q_H$ and $q_K$ directions is sharply different. Both magnetic Bragg peaks are broader than the nuclear Bragg (2, 0, 0) reflection in the reciprocal space as shown in \textbf{Figure~\ref{Fig5}B}, which indicates that the observed magnetic Bragg peaks are beyond the instrument resolution. Therefore, \textbf{Figure~\ref{Fig5}B} shows a real in-plane magnetic anisotropy.

\section{IV. Discussion}

\begin{table*}[!htdp]
\caption{Refined structural parameters (lattice constants, atomic positions, Debye-Waller factor \emph{B}, bond angles, and bond lengths), magnetic moment $\hat{\mu}$, and the corresponding goodness of refinement by the Fullprof Suite \cite{Rodriguez-Carvajal1993} from the NPD data measured at 0.5, 10 and 60 K using SPODI (FRM-II). The calculated average bond-lengths $\langle$Tb1-O1,2,3$\rangle$ and $\langle$Tb2-O1,3,4$\rangle$ and the extracted octahedral distortion parameter $\Delta$ are also listed. All atoms reside in the Wyckoff site 4c, i.e., (\emph{x}, \emph{y}, 0.25). Number in parenthesis is the estimated standard deviation of the last significant digit.}
\label{Tab:01}
{\renewcommand\tabcolsep{2.5pt}
\begin{tabular} {rccccccccc}
\hline
\hline
\multicolumn{10}{c} {Pulverized SrTb$_2$O$_4$ single crystal (Orthorhombic, space group \emph{Pnam}, $Z = 4$)}  \\*
\hline
\emph{T} (K)                       & \multicolumn{3}{c} {0.5} & \multicolumn{3}{c} {10} & \multicolumn{3}{c} {60} \\*
\hline
$a, b, c$ ({\AA}) & 10.0842(1) & 11.9920(2) & 3.4523(1) & 10.0844(1) & 11.9918(1) & 3.4522(1) & 10.0852(1) & 11.9922(1) & 3.4525(1) \\*
Atom & \emph{x} & \emph{y} &\emph{B} ({\AA}$^2$)& \emph{x} & \emph{y} &\emph{B} ({\AA}$^2$)& \emph{x} & \emph{y} &\emph{B} ({\AA}$^2$)\\*
Sr                & 0.7497(1) & 0.6487(1) & 0.85(5) & 0.7493(2) & 0.6485(2) & 0.87(7) & 0.7498(3) & 0.6491(3) & 1.04(8) \\*
Tb1               & 0.4243(2) & 0.1126(1) & 0.27(4) & 0.4241(2) & 0.1124(2) & 0.25(5) & 0.4250(3) & 0.1123(2) & 0.56(6) \\*
Tb2               & 0.4182(2) & 0.6116(1) & 0.47(4) & 0.4180(2) & 0.6116(2) & 0.50(5) & 0.4178(3) & 0.6114(2) & 0.53(7) \\*
O1                & 0.2133(2) & 0.1799(1) & 0.64(5) & 0.2138(3) & 0.1798(2) & 0.75(7) & 0.2125(3) & 0.1796(2) & 0.63(9) \\*
O2                & 0.1293(2) & 0.4818(1) & 0.16(5) & 0.1295(2) & 0.4819(2) & 0.21(7) & 0.1288(3) & 0.4824(2) & 0.43(9) \\*
O3                & 0.5092(2) & 0.7859(2) & 0.56(4) & 0.5095(2) & 0.7857(2) & 0.48(6) & 0.5095(3) & 0.7859(3) & 0.69(8) \\*
O4                & 0.4273(2) & 0.4216(1) & 0.55(5) & 0.4271(3) & 0.4218(2) & 0.49(7) & 0.4270(4) & 0.4217(2) & 0.74(9) \\*
$\hat{\mu}$(Tb1) $(\mu_\texttt{B})$   & \multicolumn{3}{c} {\emph{b}-axis: +1.88(8), \emph{c}-axis: +0.40(23)} & & & & & & \\*
\hline
$\angle$Tb1-O2-Tb1 ($^\circ$) & \multicolumn{3}{c} {92.3(1), 96.7(1)} & \multicolumn{3}{c} {92.4(1), 96.6(1)} & \multicolumn{3}{c} {92.6(1), 96.7(2)} \\*
$\angle$Tb1-O3-Tb1 ($^\circ$) & \multicolumn{3}{c} {102.3(1)} & \multicolumn{3}{c} {102.2(1)} &\multicolumn{3}{c} {102.4(1)} \\*
$\angle$Tb1-O1-Tb2 ($^\circ$) & \multicolumn{3}{c} {114.1(1)} & \multicolumn{3}{c} {114.1(2)} &\multicolumn{3}{c} {114.0(2)} \\*
$\angle$Tb1-O3-Tb2 ($^\circ$) & \multicolumn{3}{c} {128.7(1)} & \multicolumn{3}{c} {128.8(2)} &\multicolumn{3}{c} {128.7(2)} \\*
$\angle$Tb2-O1-Tb2 ($^\circ$) & \multicolumn{3}{c} {95.8(1)} & \multicolumn{3}{c} {95.8(1)} &\multicolumn{3}{c} {96.2(1)} \\*
$\angle$Tb2-O4-Tb2 ($^\circ$) & \multicolumn{3}{c} {94.1(1), 101.3(1)} & \multicolumn{3}{c} {93.9(1), 101.3(1)} &\multicolumn{3}{c} {93.8(1), 101.2(2)} \\*
$\angle$O1-Tb1-O2 ($^\circ$) & \multicolumn{3}{c} {172.1(2), 91.2(1)} & \multicolumn{3}{c} {172.3(2), 91.3(1)} &\multicolumn{3}{c} {171.7(2), 91.0(2)} \\*
$\angle$O2-Tb1-O3 ($^\circ$) & \multicolumn{3}{c} {171.8(1), 89.9(1)} & \multicolumn{3}{c} {172.0(1), 89.9(1)} &\multicolumn{3}{c} {172.2(2), 90.3(2)} \\*
$\angle$O3-Tb2-O4 ($^\circ$) & \multicolumn{3}{c} {154.0(1), 83.7(1)} & \multicolumn{3}{c} {153.9(2), 83.6(1)} &\multicolumn{3}{c} {153.8(3), 83.5(2)} \\*
$\angle$O1-Tb2-O4 ($^\circ$) & \multicolumn{3}{c} {168.3(1), 112.0(1)} & \multicolumn{3}{c} {168.3(2), 112.0(2)} &\multicolumn{3}{c} {168.0(2), 112.1(2)} \\*
\hline
$\langle$Tb1-O1,2,3$\rangle$ ({\AA}) & \multicolumn{3}{c} {2.3088(8)} & \multicolumn{3}{c} {2.3084(10)} &\multicolumn{3}{c} {2.3073(13)} \\*
$\langle$Tb2-O1,3,4$\rangle$ ({\AA}) & \multicolumn{3}{c} {2.3220(8)} & \multicolumn{3}{c} {2.3232(11)} &\multicolumn{3}{c} {2.3216(14)} \\*
$\Delta$ ($\times10^{-4}$)    & \multicolumn{3}{c} {Tb1: 10.905, Tb2: 1.894} & \multicolumn{3}{c} {Tb1: 10.729, Tb2: 2.077} &\multicolumn{3}{c} {Tb1: 9.987, Tb2: 2.079} \\*
\hline
$R_p, R_{wp}, R_{exp}, \chi^2$ & \multicolumn{3}{c} {1.73, 2.28, 1.46, 2.44} & \multicolumn{3}{c} {2.38, 3.06, 3.08, 0.986} &\multicolumn{3}{c} {2.83, 3.65, 3.95, 0.856} \\*
\hline
\hline
\end{tabular}}{}
\end{table*}

To quantitatively estimate the in-plane anisotropy, we take the FWHM of the nuclear Bragg (2, 0, 0) peak as the detecting accuracy which is convoluted in fitting the magnetic peaks by a Gaussian function shown as the solid lines in \textbf{Figure~\ref{Fig5}B}. This results in FWHM = 0.0183(1) and 0.0492(2) {\AA}$^{-1}$ along the $q_H$ and $q_K$ directions, respectively, implying highly anisotropic in-plane spin correlations consistent with the observation that strong magnetic frustration exists in SrTb$_2$O$_4$. We roughly estimate the spin-correlation length ($\xi$) by $\xi = \frac{2\pi}{{\texttt{FWHM}}}$, i.e., $\xi_H = 343.7(22)$ {\AA} and $\xi_K = 127.6(4)$ {\AA}. Therefore, $\frac{\xi_H}{\xi_K} = 2.69(2)$. Similar in-plane anisotropic magnetic correlations were also observed in the iron-based superconductors \cite{Li2010, Li2010-1, Proke2012, Li2011, Xiao2013} that are highly frustrated, too, where its microscopic origin, from the ellipticity of the electron pockets or the competing exchange interactions associated with the local-moment magnetism, is still being strongly argued \cite{Zhao2009, Diallo2010, Shen2011, Wang2013}. It is undoubted that the observed in-plane magnetic anisotropy in SrTb$_2$O$_4$ indicates an appearance of the competing spin exchanges and is certainly associated with a description of the purely-localized magnetism of ionic Tb$^{3+}$ ions. A deeper understanding of the insulating state necessitates theoretical band structure calculations. We tentatively estimate the compatibility between ordered magnetic and nuclear crystalline domains based on the non-deconvoluted FWHM ($\kappa$) of the Bragg (1.6, 1, 0) ($\kappa_m$ = 0.0300(7) {\AA}$^{-1}$) and (2, 0, 0) ($\kappa_n$ = 0.0238(2) {\AA}$^{-1}$) peaks, i.e., $\kappa_n$/$\kappa_m$ = 79(2)\%, which implies that the incommensurate AFM structure orders with a long-range fashion relative to the underlying lattice of the single crystal.

We further analyze the spin-correlation length with our NPD data (\textbf{Figure~\ref{Fig2}C}). Firstly, it is pointed out that the positive and negative momenta cannot technically be differentiated in a NPD study. As shown in \textbf{Figure~\ref{Fig6}A}, taking into account the corresponding SPODI instrument resolution (dashed line) \cite{Hoelzel2012}, a Gaussian fit (solid line) to the AFM Bragg (1.5924, 1.0059, 0) peak (squares) results in an average $\xi_{\texttt{AFM}} = 864(36)$ {\AA} in real space. This indicates that the AFM ordering observed in SrTb$_2$O$_4$ is indeed of long range in character in comparison with the reported extremely-broad magnetic diffuse scattering which was attributed to the presence of short-ranged magnetic ordering in polycrystalline Sr$RE_2$O$_4$ (\emph{RE} = Ho, Er, Dy) samples in the study of reference \cite{Karunadasa2005}. With the same method utilized in the analysis of the data as shown in \textbf{Figure~\ref{Fig6}A}, we also analyze the NPD peak of the nuclear Bragg (2, 0, 0) reflection as shown in \textbf{Figure~\ref{Fig6}B} and extract that $\xi_{\texttt{(200)}} = 1304(34)$ {\AA}. This indicates that $\xi_{\texttt{AFM}}/\xi_{\texttt{(200)}}$ = 66(3)\% basically in accord with the compatibility between ordered magnetic and nuclear crystalline domains extracted with our single-crystal neutron-scattering data. Since our NPD data were collected from a pulverized SrTb$_2$O$_4$ single crystal, that $\xi_{\texttt{AFM}}$ is $\sim$ 2.5 times larger than $\xi_H$ may indicate that there have strong magnetic and crystalline domain effects in single-crystal SrTb$_2$O$_4$, or a large part of spins are blocked probably due to a pining effect by strains accumulated during single crystal growth. In any case, this difference between single-crystalline and polycrystalline samples in turn supports the fact that there is a strong magnetic frustration in single-crystal SrTb$_2$O$_4$. Further studies with high pressures would be of great interest.

In most cases, the strength of the indirect magnetic interactions such as conventional double- or super-exchange \cite{Santen1950} can be influenced more or less by the value of the revelent bond angle \cite{Li2007-1, Li2007-2, Li2009, Ma2013}, e.g. the $\angle$Tb-O-Tb bond angles in SrTb$_2$O$_4$ as listed in \textbf{Table 1} (see also \textbf{Figure~\ref{Fig7}}). However, the respective values of $\angle$Tb-O-Tb display no appreciable difference within accuracy between 0.5 and 10 K (\textbf{Table 1}), below and above the $T_{\texttt{N}}$, respectively, which may indicate an invalidity of the two conventional magnetic coupling mechanisms (double- or super-exchange) in SrTb$_2$O$_4$. This is consistent with the study of SrTm$_2$O$_4$ \cite{Li2014-1} and in excellent agreement with our transport study, where any attempts to measure possible resistivity in SrTb$_2$O$_4$ from 2 to 300 K were fruitless. We estimate that the resistance of the single crystal measured is beyond at least 10$^6$ ohm. We thus conclude that SrTb$_2$O$_4$ is a robust insulator, and the electrons responsible for the incommensurable antiferromagnetism are mainly from the localized 4$f^8$ shell of the ionic Tb$^{3+}$ ions. In this localized picture, the interionic exchange interactions dominate for the formation of the magnetic structure \cite{Diep2004, Li2012}. The nearest Tb neighbours are stacked linearly along the \emph{c} axis (\textbf{Figure~\ref{Fig3}B}). The shortest Tb1-Tb1 and Tb2-Tb2 have the same bond length. However, the NN Tb1 ions have a ferromagnetic (FM) arrangement. By contrast, the interaction between the NN Tb2 ions is blocked unexpectedly (\textbf{Figure~\ref{Fig3}B}). There is no appreciable difference in the NN Tb-Tb bond length, i.e., the \emph{c} lattice constant, between 0.5 and 10 K (\textbf{Table 1}), which probably rules out the potential direct exchange interaction consistent with the fact that unpaired 4\emph{f} electrons are deeply embedded under the $5s^2p^6$ shells and also indicates that the prevalent dipole-dipole interaction is subjected to some condition, i.e., the octahedral distortion as discussed below, in agreement with the study of SrTm$_2$O$_4$ \cite{Li2014-1}.

As a non-Kramers ion, Tb$^{3+}$ ($S = 3, L = 3, J = 6, g_J = 1.5$) in principle keeps the time reversal symmetry and doesn't show any energy degeneracy in the presence of the purely-localized electric field. However, we refine two kinds of octahedra as shown in \textbf{Figure~\ref{Fig7}}: Tb1O$_6$ and Tb2O$_6$, corresponding to the partially-ordered and totally-frozen Tb1 and Tb2 ions, respectively. The average octahedral distortion \cite{Li2007-1, Li2007-2} can be quantitatively measured by the parameter $\Delta$ defined as:
$\Delta = \frac{1}{6} \sum\limits_{n = 1}^{6} \mathlarger{\mathlarger{[}}\frac{(d_n - \langle d \rangle)}{\langle d \rangle}\mathlarger{\mathlarger{]}}^2$,
where $d_n$ and $\langle d \rangle$ are the six Tb-O bond lengths along the six crossed directions (\textbf{Figure~\ref{Fig7}}) and the mean Tb-O bond length (\textbf{Table 1}), respectively. It is noteworthy that the $\Delta$ values of the Tb1 and Tb2 ions are in the same magnitudes as those of the Mn$^{3+}$ Kramers and Mn$^{4+}$ non-Kramers ions, respectively, in the Jahn-Teller (JT) distorted regime of single-crystal La$_{\frac{7}{8}}$Sr$_{\frac{1}{8}}$MnO$_3$ \cite{Li2009}. This sharp contrast implies that the Tb1 ions are strongly distorted, while the Tb2 ions behave normally within the non-Kramers scheme. Therefore, the $\Delta$ magnitude that reflects the ion local symmetry and thus the strength of the surrounding CEF directly determines the existence of the magnetic ordering, which is supported by the observation that below $T_\texttt{N}$ the respective $\Delta$ values of the Tb1 and Tb2 ions change oppositely with temperature (\textbf{Figure~\ref{Fig8}}). We therefore infer that one possible reason for the formation of the incommensurable magnetic structure is the modulated distribution of the $4f^16s^2$ valence electrons which modify the surrounding environment experienced by the localized unpaired 4\emph{f} electrons. The corresponding modulation of the local symmetry may plausibly be attributed to the spatial zigzag-type Tb arrangements along the \emph{a} and \emph{b} axes in the process of forming the crystallographic domains. This is supported by the fact that the honeycomb columns run straightly along the \emph{c} axis, and there is no spin modulation at all in that direction.

Based on the refined Tb-O bond lengths, we deduce two distortion modes for the Tb1O$_6$ and Tb2O$_6$ octahedra (\textbf{Figure~\ref{Fig7}}), respectively. The possible product of the Tb1 subjected stress-vectors (small arrows) should point qualitatively to the direction of the Tb1 moment, implying a strong single-ion anisotropy. This JT-like distortion mode leads to the large $\Delta$ value of the Tb1 ions, and possibly lifts further the degenerate multiplets. By contrast, the Tb2 ions are subjected to opposing stresses in all the three pair-directions. In this case, the octahedral distortion strongly depends on their competing strengths. This mode makes the small $\Delta$ value of the Tb2 ions and their potential total magnetic moments quenched vitally.

The maximum Tb1 moment size is mere 1.92(6) $\mu_\texttt{B}$, 21.3(7)\% of the theoretical saturation value ($g_J J = $ 9 $\mu_\texttt{B}$). It is of particular interest to explore the frustrating mechanism. The virtual non-Kramers state of the Tb2 site reduces the total moment size per molar formula by 50\%. The Tb1 moment fluctuates like a wave defined as $\hat{\mu}$ = $|\hat{\mu}_{\texttt{max}}| cos(\texttt{Q}_{\texttt{AFM}} \cdot R_x + \phi)$, where $R_x$ is a spin coordinate along the \emph{a} axis and $\phi$ is a phase parameter. The existence of the strong single-ion anisotropy indicates a large CEF effect which should be comparable to the energy scale of the magnetic interactions. We have shown the clear evidence for a large magnetic exchange anisotropy (\textbf{Figure~\ref{Fig5}}), which is ascribed to the anisotropic dipole-dipole interaction. The NN magnetic arrangement is FM (\textbf{Figure~\ref{Fig3}B}), implying no possibility for a magnetic frustration. The NNN magnetic configurations display a dual character, i.e., FM and AFM for the equivalent Tb11 and Tb12 sites, respectively. This sharp difference may frustrate the Heisenberg-exchange coupled NNN spins.

\section{V. Conclusions and outlook}

To summarize, we have synthesized large enough SrTb$_2$O$_4$ single crystals suitable for neutron scattering studies and revealed a modulated spin structure in SrTb$_2$O$_4$ with the highest AFM transition temperature at $T_\texttt{N} =$ 4.28(2) K in the Sr$RE_2$O$_4$ family, which provides a technically friendly platform to explore the related magnetic coupling and frustrating mechanisms. Our studies show that the localized Tb1 moments lie in the \emph{bc} plane with the FM chains along both the \emph{b} and \emph{c} directions and the AFM modulation mainly along the \emph{a} axis. We have found two distinct octahedra for the non-Kramers Tb$^{3+}$ ions: Tb1O$_6$ being strongly distorted, corresponding to the partially-ordered moments; Tb2O$_6$ being frustrated entirely in the non-Kramers state. Therefore, the octahedral distortion has a decisive influence on the Hund's rule magnetic ground state $(^7F_6)$ and the related frustrations. The magnetocrystalline anisotropy is crucial in determining the direction of the ordered moments. The direct NN interaction results in a FM arrangement for the Tb1 ions along the \emph{c} axis, and the different NNN Tb configurations (FM and AFM) further lift the magnetic frustration. The present results make SrTb$_2$O$_4$ a particularly significant compound in the family for theoretical and further experimental studies. Inelastic neutron-scattering studies to determine the detailed crystal-field and magnetic-interaction parameters would be of great interest. The factors that influence the value of the AFM transition temperature would be further explored in combination with theoretical calculations.

\section{Acknowledgements}

This work at RWTH Aachen University and J$\ddot{\texttt{u}}$lich Centre for Neutron Science JCNS Outstation at ILL was funded by the BMBF under contract No. 05K10PA3. H.F.L thanks the sample environment teams at ILL and FRM-II for expert technical assistances.

\section{Author contributions}

C.Z, P.M, and H.F.L prepared the polycrystals and grew the single crystals.
A.S and H.F.L performed the SPODI experiments and analyzed the data.
M.B, B.Y.H, and H.F.L performed the IN3 experiments.
A.W and H.F.L performed the D7 experiments and analyzed the data.
K.S, W.S, and H.F.L performed the IN12 experiments and analyzed the data.
W.S, K.S, E.R, and H.F.L performed the D23 experiments.
H.F.L, C.Z, A.S, A.W, K.S, W.S, M.B, E.R, B.Y.H, P.M, G.R, and Th.B discussed and analyzed the results.
H.F.L wrote the main manuscript text.
C.Z, A.S, A.W, K.S, B.Y.H, G.R, and Th.B commented on the manuscript and all authors reviewed the paper.
H.F.L conceived and directed the project.

\section{Conflict of Interest Statement}

The authors declare that the research was conducted in the absence of any commercial or financial relationships that could be construed as a potential conflict of interest.


\end{document}